\documentclass[twocolumn]{aastex63}

\received{June 1, 2019}
\revised{January 10, 2019}
\accepted{\today}
\submitjournal{AJ}

\shorttitle{Additional Calibration of the Ultra-Violet Imaging Telescope on board AstroSat}
\shortauthors{Tandon et al.}

\begin{document}

\title{Additional Calibration of the Ultra-Violet Imaging Telescope on board AstroSat}

\correspondingauthor{C. S. Stalin}
\email{stalin@iiap.res.in}

\author{S.N. Tandon}
\affiliation{Inter-University Center for Astronomy and Astrophysics, Pune 411007, India}

\author{J. Postma}
\affiliation{.  University of  Calgary, 2500 University Drive NW, Calgary, Alberta Canada}

\author{P. Joseph}
\affiliation{Indian Institute of Astrophysics, Koramangala II Block, Bangalore 560034, India}

\author{A. Devaraj}
\affiliation{Indian Institute of Astrophysics, Koramangala II Block, Bangalore 560034, India}

\author{A Subramaniam}
\affiliation{Indian Institute of Astrophysics, Koramangala II Block, Bangalore 560034, India}

\author{I. V. Barve}
\affiliation{Indian Institute of Astrophysics, Koramangala II Block, Bangalore 560034, India}

\author{K. George}
\affiliation{Indian Institute of Astrophysics, Koramangala II Block, Bangalore 560034, India}

\author{S. K. Ghosh}
\affiliation{Tata Institute of Fundamental Research, Mumbai, India}

\author{V. Girish}
\affiliation{ISRO Headquarters, Bengaluru, India}

\author{J. B. Hutchings}
\affiliation{National Research Council of Canada, Herzberg Astronomy and Astrophysics, 5071 West Saanich Road, Victoria, BC V9E 2E7, Canada}

\author{P. U. Kamath}
\affiliation{Indian Institute of Astrophysics, Koramangala II Block, Bangalore 560034, India}

\author{S. Kathiravan}
\affiliation{Indian Institute of Astrophysics, Koramangala II Block, Bangalore 560034, India}

\author{A. Kumar}
\affiliation{Indian Institute of Astrophysics, Koramangala II Block, Bangalore 560034, India}

\author{J. P. Lancelot}
\affiliation{Indian Institute of Astrophysics, Koramangala II Block, Bangalore 560034, India}

\author{D. Leahy}
\affiliation{University of Calgary, 2500 University Drive NW, Calgary, Alberta, Canada}

\author{P.K. Mahesh}
\affiliation{Indian Institute of Astrophysics, Koramangala II Block, Bangalore 560034, India}

\author{R. Mohan}
\affiliation{Indian Institute of Astrophysics, Koramangala II Block, Bangalore 560034, India}

\author{S. Nagabhushana}
\affiliation{Indian Institute of Astrophysics, Koramangala II Block, Bangalore 560034, India}

\author{A. K. Pati}
\affiliation{Indian Institute of Astrophysics, Koramangala II Block, Bangalore 560034, India}

\author{N. Kameswara Rao}
\affiliation{Indian Institute of Astrophysics, Koramangala II Block, Bangalore 560034, India}

\author{K. Sankarasubramanian}
\affiliation{U. R. Rao Satellite Centre, HAL Airport Road, Bengalure 560 017, India}

\author{S. Sriram}
\affiliation{Indian Institute of Astrophysics, Koramangala II Block, Bangalore 560034, India}

\author{C. S. Stalin}
\affiliation{Indian Institute of Astrophysics, Koramangala II Block, Bangalore 560034, India}

\begin{abstract}

Results of the initial calibration of the Ultra-Violet Imaging Telescope (UVIT) were 
reported 
earlier by \cite{2017AJ....154..128T}. The results reported earlier were based on 
the ground calibration as well as the first observations in orbit. Some 
additional data from the ground calibration and data from more in-orbit 
observations have been used to improve the results. In particular, extensive 
new data from in-orbit observations have been used to obtain (a) new
photometric calibration which includes (i) zero-points (ii) 
flat fields (iii) saturation, (b) sensitivity variations (c) 
spectral calibration for the near Ultra-Violet 
(NUV; 2000$-$3000 \AA) and far Ultra-Violet (FUV; 1300$-$1800 \AA) 
gratings, (d) point spread function  and (e) astrometric calibration which includes distortion.
Data acquired over the last three years show continued
good performance of UVIT with no reduction in sensitivity in both the UV
channels.
\end{abstract}

\keywords{Ultraviolet astronomy (1736) -- Ultraviolet telescopes (1743) --- Astronomical instrumentation (799) }

\section{Introduction} \label{sec:intro}

Ultra-Violet Imaging Telescope (UVIT) is one of the five payloads on board the 
Indian multi-wavelength astronomy satellite AstroSat \citep{2006AdSpR..38.2989A}. Four of the five instruments 
on AstroSat observe in the soft and hard X-ray bands, while UVIT observes in 
the Ultra-Violet bands. 
The primary aim of UVIT is simultaneous imaging in the far Ultra-Violet (FUV; 1300$-$1800 \AA) 
and the near Ultra-Violet (NUV; 2000$-$3000 \AA) channels over a field of $\sim$28$^{\prime}$ diameter with a spatial 
resolution $<$ 1.5$^{\prime\prime}$.  For both FUV and NUV channels, multiple filters 
are provided for observations in a narrower band, and options for slit-less 
spectroscopy, with a resolution of $\sim$80, too are provided. The 
initial calibration of UVIT was reported in \citealt{2017AJ....154..128T},
(hereinafter referred to as Paper-1), which were 
based on the measurements done on ground and initial observations on the sky. 
With accumulation of in-orbit observations carried out over about 30 months on 
the calibration fields, the calibration has been refined and the results are reported here.  
These results supersede the ones reported in Paper-1.
The paper is organised as follows: important details of the instrument are 
described in Section 2, a brief description of the calibration reported 
in Paper-1 is presented in Section 3, details of the 
new  observations are given in Section 4, results of the calibration are given in Section 5 
and Section 6 is devoted to conclusions.  

\begin{deluxetable*}{llll}
\tablecaption{Key parameters of the three channels of UVIT \label{table-1}}
\tablewidth{0pt}
\tablehead{
\colhead{Parameter}  & \colhead{FUV}  &  \colhead{NUV}  &  \colhead{VIS$^{a}$}  
  }
%\decimalcolnumbers
\startdata
Wavelength (\AA)                  &  1300 $-$ 1800    & 2000 $-$ 3000   & 3200 $-$ 5500  \\
Mean Wavelength$^b$ (\AA)             & 1481            & 2418          & 4200         \\
Mean Effective Area (cm$^2$)      & $\sim$10        &  $\sim$40     &  $\sim$50         \\
Field of view (diameter - arcmin) & 28              &  28           & 28           \\
Plate Scale ($\prime\prime$/pixel)             & 3.33            & 3.33          & 3.30         \\
Astrometric accuracy (rms)        & 0.4$^{\prime\prime}$             & 0.4$^{\prime\prime}$           &              \\
Zero-point magnitude$^c$              & 18.1            & 19.8          &              \\
Spatial resolution$^d$ (FWHM)         & 1.3$^{\prime\prime}$ $-$ 1.5$^{\prime\prime}$     & 1.2$^{\prime\prime}$ $-$ 1.4$^{\prime\prime}$   & 2.5$^{\prime\prime}$          \\
Spectral resolution$^e$ (\AA)         & 17              & 33            &              \\
Saturation (counts/sec)$^f$ (10\%)    & 6               & 6             &              \\ \hline
\enddata
\tablecomments{
$^a$ For the VIS channel all the parameters are based on ground calibration. This 
channel is operated in integration mode. Photometric calibration is not done as 
we don't expect doing science with VIS channel observations. This channel is 
meant for aspect correction. \\
$^b$ The mean wavelength is for the filter with maximum bandwidth, and is 
obtained by weighting wavelengths with the corresponding effective area as 
measured in calibration on the ground.  \\
$^{c}$ The zero point magnitude (for the filter with maximum bandwidth) is in 
AB system and refers to the average flux of HZ4 in the band \\
$^d$ It depends on perturbations in the pointing. \\
$^e$ These are for the gratings. \\
$^f$  The saturation is given for the full field images. These are taken at a 
rate 28.7 frames/s; images for partial field are taken at higher frequency of 
the frames and the range of linearity is higher }
\end{deluxetable*}

\section{Instrumentation}
Two co-aligned Ritchey-Chretien telescopes, each of aperture  $\sim$375 mm, 
are used to image in FUV (1300$-$1800 \AA), NUV (2000$-$3000 \AA) and 
VIS (3200$-$5500 \AA) channels. As drift in the pointing can be many arc-seconds 
at rates up to 2 arc-second/sec, images are generated by combining 
short-exposure frames   ($<<$ 1 sec) after applying corrections for drift.
%using shift and add algorithm. 
Many of 
the fields would not have enough flux in NUV/FUV for tracking the drift, 
therefore images taken with the VIS channel, at regular intervals of $\sim$1 sec, 
are used to monitor the drift. Each of the three channels has a filter wheel to 
select a narrower band for imaging. For low resolution slit-less spectroscopy, 
one grating is provided in the NUV wheel and two gratings (with orthogonal 
dispersions) are provided in the FUV wheel. Intensified CMOS imagers, of aperture 
39 mm, are used for all the three channels. The imagers can work either in 
photon-counting mode (with high intensification/high electron multiplication due to high voltage in the micro channel plate; MCP) or in integration mode (with 
low intensification/low electron multiplication due to low voltage in MCP). The CMOS imagers have 512 $\times$ 512 pixels, and each of these 
pixels is mapped to 8 $\times$ 8 sub-pixels in the final image to get a 
plate-scale of $\sim$0.416$^{\prime\prime}$ per sub-pixel.  Observations can be 
carried out for the full field of $\sim$28$^{\prime}$ diameter, at a rate 
$\sim$29 frames/sec or for a selectable partial field at a higher rate. Key 
performance parameters of the three channels are presented in Table \ref{table-1}, and 
properties of the filters are shown in Table \ref{table-2}. For more details of the 
instrument the reader is referred to \cite{2017JApA...38...28T} and the 
references therein.

\begin{deluxetable}{llcr}
\tablecaption{Properties of the individual filters of UVIT.  
Here, $\lambda_{mean}$ is the mean wavelength (estimated by weighting the 
wavelengths with the effective areas) and $\Delta\lambda$ is the band width 
(between the wavelengths with effective area 50\% of the peak) \label{table-2}.}
\tablewidth{0pt}
\tablehead{
\colhead{Filter Name}   & \colhead{Filter}  & \colhead{$\lambda_{mean}$ (\AA)}    & \colhead{$\Delta\lambda$ (\AA)} 
         }
%\decimalcolnumbers
\startdata
F148W	& CaF2-1	& 1481	& 500~~~   \\
F148Wa	& CaF2-2	& 1485	& 500~~~   \\
F154W	& BaF2		& 1541  & 380~~~    \\
F172M	& Silica	& 1717	& 125~~~   \\
F169M	& Sapphire	& 1608	& 290~~~  \\
N242W	& Silica-1	& 2418	& 785~~~  \\
N242Wa	& Silica-2	& 2418	& 785~~~  \\
N245M	& NUVB13	& 2447	& 280~~~  \\
N263M	& NUVB4		& 2632  & 275~~~  \\
N219M	& NUVB15	& 2196	& 270~~~  \\
N279N	& NUVN2		& 2792  & 90~~~   \\ 
V347M$^{\dagger}$       & VIS1          & 3466  & 400~~~ \\
V391M   & VIS2          & 3909  & 400~~~ \\
V461W   & VIS3          & 4614  & 1300~~~ \\
V420W   & BK7           & 4200  & 2200~~~ \\
V435ND  & ND1           & 4354  & 2200~~~ \\
\enddata  
\tablecomments{
$^\dagger$ The VIS channel is only meant for aspect correction and not expected 
for science observation.   The inbuilt safely features of UVIT require the 
count rates in VIS channel not to exceed 4800 c/s for any individual point source. To enable 
observations of fields that have optical sources with varied brightness levels, 
the VIS channel too have different filters such as V347M, V391M, V461W, V420W and V435ND.}  

\end{deluxetable}

\section{Calibration required}
All the calibration can be divided into four sets (see 
Paper-1 for details). The scope of these is 
briefly described below
\begin{enumerate}
\item Photometric calibration: This includes (a) zero point magnitudes at the centre of 
the field for different filters obtained from the observations on a 
standard  star (HZ 4 is observed for this), (b) flat-field variations remaining after correcting 
for variations in the sensitivity of the detectors as observed in the ground 
calibration at the mean wavelengths, and (c) correction for saturation in the 
photon-counting mode. 
\item  Monitoring sensitivity of FUV and NUV channels: As throughput of the 
FUV and NUV channels can be reduced by depositions of contaminants on the optics, 
as well as due to ageing of the filters, the MCPs and coatings on the mirrors,
sensitivity of both the channels is monitored every few months.

%Monitoring sensitivity of FUV and NUV channels: As transmission of the FUV and NUV channels 
%can be severely reduced by monomolecular depositions of contaminants on the optics, 
%the sensitivity of both the channels is monitored every few months.  
\item Spectral calibration for the NUV and FUV gratings: This includes (a) effective 
area as a function of wavelength, (b) dispersion, and (c) resolution.
\item Point Spread Function, including encircled energy as a function of radius.
\item Astrometric calibration giving estimates of errors in the positions after 
correcting for distortions in the detectors as per the ground calibration.
\end{enumerate}

%The results of these calibration were earlier reported 
%in Paper-1.
%With the availability of more data from in-orbit observations, improved results, 
%particularly on the flat-field variations have been obtained. All the results of 
%calibration, either in original or as modified by additional calibration, 
%are presented in the next section. 
%In the following we shall refer to the 
%results reported by \cite{2017AJ....154..128T} as “reported earlier”.
The results of calibration were earlier reported
in Paper-1. With the availability of more data from in orbit observations, 
improved results, particularly on the flat field variations have been obtained, 
and are presented in Section 5.

\section{Observations and Data Analysis}
The results reported here are based on the observations of HZ 4, three overlapping fields in 
the Small Magellanic Cloud (SMC) and NGC 188. All the observations were made in photon-counting 
mode of the detectors. Most of the observations were made with full field, but 
some of the observations for HZ 4 were made with partial field to get frame rates up 
to $\sim$292 frames/sec to minimise the effect of saturation. More details for the 
observations can be found in Paper-1. To monitor any possible 
reduction in throughput of FUV and NUV channels, due to deposition of any contaminants 
on the optics or due to ageing of the filters and the MCPs, 
a field in NGC 188 was selected as its high declination provides visibility 
throughout the year.  The three fields in SMC were selected as follows: the first 
field was selected away from the centre of SMC to avoid the brightest part, the 
other two fields were selected to get shifts of $\sim$6$^{\prime}$ in two orthogonal 
directions with reference to the first field. The shifts of $\sim$6$^{\prime}$ 
were used to get a good overlap between the fields as well as obtain a separation 
of several arcminutes for positions of common objects in the three images. The 
differences in count-rates between the three positions provide data on differential 
sensitivity across more than one thousand separation-vectors distributed over the area of 
the detector.  All the images were generated with
CCDLAB \citep{2017PASP..129k5002P}. In CCDLAB\footnote{https://sourceforge.net/projects/ccdlab/}, 
each event is corrected for position 
as well as flat-field.  The positions are corrected for: (i) a bias 
called fixed pattern noise, (ii) distortion, and  (iii) drift of the pointing. The 
correction for flat-field is only for the spatial variations in sensitivity of  
the detectors as measured during pre-launch calibration,  i.e. possible contributions of other
optical elements are not included and are to be deduced from these images as 
discussed under "Remainders of flat field" in Section 5.1.3.
%i.e. the variations caused by other optical 
%elements are not considered (see Section 4.2, point-3) 
 The corrections for distortion were 
generated by a reanalysis of all data from ground calibration done at the 
University of Calgary and at the Indian Institute of Astrophysics (IIA), and 
have small differences compared to what were used for the results reported 
in Paper-1. The arrays for flat-field corresponding to the pre-launch 
calibration, too were  regenerated by correcting the position 
of each event for distortion on the detector, which is equivalent to correcting 
for errors due to variations in plate-scale caused by distortions, and have 
small differences compared to what were used for the results reported 
in Paper-1.      
Filters F148Wa and N242Wa are spare filters which were not used for any observations 
and hence no results are presented for these. 

\begin{deluxetable}{lllrlr}
\tablecaption{Normalised counts/sec and zero point magnitudes for the different UVIT filters \label{table-3}}
\tablewidth{0pt}
\tablehead{
\colhead{Name}   & \colhead{Filter}   & \colhead{$\lambda_{mean}$ \AA}   & \colhead{Normalised}   & \multicolumn{2}{c} {ZP magnitude}  \\ 
\colhead{}       & \colhead{}         & \colhead{}                       & \colhead{c/s for HZ 4}  & \colhead{Value}  & \colhead{error}
}
%\decimalcolnumbers
\startdata
F148W	& CaF2-1	& 1481	& 23.52	 & 18.097 &	0.010   \\
F154W	& BaF2	        & 1541	& 20.68	 & 17.771 &	0.010   \\
F169M	& Sapphire	& 1608	& 16.16	 & 17.410 &	0.010   \\
F172M	& Silica	& 1717	& 5.460	 & 16.274 &	0.020   \\
N242W	& Silica-1	& 2418	& 127.800 & 19.763 &	0.002  \\
N219M	& NUVB15	& 2196	& 7.360	 & 16.654 &	0.020   \\
N245M	& NUVB13	& 2447	& 36.97	 & 18.452 &	0.005  \\
N263M	& NUVB4	        & 2632	& 27.16	 & 18.146 &	0.010   \\
N279N	& NUVN2	        & 2792	& 5.37	 & 16.416 &	0.010   \\ \hline
\enddata
\end{deluxetable}

\begin{deluxetable*}{llllllllllllllllll}
\tablecaption{Results of ground calibration on effective areas of the filters as a 
function of wavelength for the centre of the field. 
Wavelength($\lambda$) is in \AA ~and effective area (E) is in 
square cm \label{table-4}.}
\tablewidth{0pt}
\tablehead{
\multicolumn{2}{c}{F148W} & \multicolumn{2}{c}{F154W} &\multicolumn{2}{c}{F169M} &\multicolumn{2}{c}{F172M} &\multicolumn{2}{c}{N242W} & \multicolumn{2}{c}{N219M} & \multicolumn{2}{c}{N245M} & \multicolumn{2}{c}{N263M} & \multicolumn{2}{c}{N279N} \\
$\lambda$  & E      & $\lambda$      &  E    & $\lambda$   &  E     & $\lambda$   & E & $\lambda$   & E  & $\lambda$   & E  & $\lambda$   & E  & $\lambda$   & E & $\lambda$   & 
E 
}  
%\decimalcolnumbers
\startdata
1250       & 0      &  1340 &  0.00  &  1420  & 0.0        & 1620   & 0.70 & 1700 & 4.08  & 1937  & 0.01 & 2148 & 0.39 & 2462 & 0.11 & 2705 & 0.10 \\
1270       & 12.11  &  1360 &  4.89  &  1440  & 0.28       & 1650   & 3.65 & 1750 & 2.36  & 2000  & 0.80 & 2191 & 0.40 & 2496 & 12.12 & 2712 & 0.19  \\
1300       & 12.11  &  1400 & 13.15  &  1480  & 11.25      & 1670   & 6.62 & 1823 & 0.68  & 2001  & 1.56 & 2209 & 1.21 & 2497 & 10.15 & 2719 & 0.17 \\
1360       & 11.88  &  1440 & 11.95  &  1540  & 11.69      & 1700   & 8.62 & 1879 & 1.28  & 2009  & 2.04 & 2221 & 2.23 & 2498 & 13.80 & 2728 & 0.47 \\
1400       & 11.98  &  1480 & 11.69  &  1600  & 10.06      & 1720   & 9.57 & 1937 & 1.97  & 2019  & 2.40 & 2235 & 3.44 & 2499 & 15.49 & 2733 & 0.98 \\
1440       & 10.36  &  1540 & 12.44  &  1650  & 10.17      & 1750   & 7.02 & 2000 & 6.41  & 2047  & 3.72 & 2275 & 14.54 & 2500 & 17.45 & 2739 & 2.53  \\
1480       & 11.70  &  1650 & 11.42  &  1700  & 9.14       & 1770   & 7.18 & 2030 & 27.65 & 2056  & 6.75 & 2282 & 16.30 & 2502 & 19.28 & 2741 & 3.74 \\
1540       & 13.29  &  1700 & 10.28  &  1750  & 6.34       & 1800   & 1.84 & 2067 & 51.97 & 2065  & 7.14 & 2289 & 17.84 & 2504 & 22.51 & 2743 & 5.47 \\
1600       & 11.61  &  1750 &  7.16  &  1780  & 2.74       & 1830   & 0.00 & 2138 & 53.61 & 2074  & 7.74 & 2294 & 19.61 & 2536 & 38.61 & 2745 & 7.46 \\
1650       & 11.78  &  1800 &  0.00  &  1800  & 0.00       &        &      & 2214 & 56.25 & 2087  & 8.14 & 2301 & 21.37 & 2551 & 40.24 & 2746 & 8.89 \\
1700       & 10.58  &       &        &        &            &        &      & 2296 & 58.01 & 2114  & 10.82 & 2308 & 23.13 & 2602 & 41.07 & 2747 & 10.40 \\
1750       & 7.37   &       &        &        &            &        &      & 2385 & 59.11 & 2124  & 11.06 & 2313 & 25.33  & 2649 & 38.28 & 2748 & 12.00 \\
1800       & 0.00   &       &        &        &            &        &      & 2461 & 56.09 & 2133  & 11.52 & 2320 & 26.66 & 2703 & 35.97 & 2749 & 14.64 \\
           &        &       &        &        &            &        &      & 2499 & 55.47 & 2147  & 11.98 & 2369 & 37.13 & 2752 & 29.31 & 2750 & 14.38 \\
           &        &       &        &        &            &        &      & 2537 & 54.76 & 2159  & 12.44 & 2381 & 39.26 & 2797 & 11.85 & 2751 & 17.06 \\
           &        &       &        &        &            &        &      & 2550 & 54.29 & 2191  & 13.72 & 2388 & 40.95 & 2798 & 10.30 & 2752 & 19.27 \\
           &        &       &        &        &            &        &      & 2600 & 51.90 & 2202  & 13.72 & 2400 & 42.65 & 2800 & 7.95  & 2754 & 21.47 \\
           &        &       &        &        &            &        &      & 2650 & 46.61 & 2216  & 12.99 & 2409 & 44.14 & 2802 & 6.18  & 2756 & 23.29 \\
           &        &       &        &        &            &        &      & 2700 & 43.07 & 2232  & 12.76 & 2440 & 44.90 & 2805 & 5.13  & 2759 & 25.02 \\
           &        &       &        &        &            &        &      & 2750 & 35.01 & 2284  & 11.79 & 2452 & 46.31 & 2846 & 0.00  & 2761 & 25.88 \\
           &        &       &        &        &            &        &      & 2800 & 30.10 & 2295  & 11.27 & 2466 & 47.72 &      &       & 2764 & 26.74 \\
           &        &       &        &        &            &        &      & 2850 & 23.32 & 2311  & 10.48 & 2485 & 48.53 &      &       & 2770 & 27.42 \\
           &        &       &        &        &            &        &      & 2900 & 16.93 & 2319  & 9.17  & 2506 & 48.72 &      &       & 2778 & 24.01 \\
           &        &       &        &        &            &        &      & 2950 & 11.36 & 2367  & 0.76  & 2520 & 49.91 &      &       & 2786 & 24.22 \\
           &        &       &        &        &            &        &      & 3000 & 7.15  & 2382  & 0.25  & 2541 & 49.63 &      &       & 2794 & 24.06 \\
           &        &       &        &        &            &        &      & 3050 & 5.04  & 2395  & 0.25  & 2560 & 49.82 &      &       & 2799 & 23.61 \\
           &        &       &        &        &            &        &      &      &       & 2410  & 0.25  & 2569 & 49.98 &      &       & 2803 & 23.08 \\
           &        &       &        &        &            &        &      &      &       &       &       & 2576 & 45.74 &      &       & 2807 & 22.48 \\
           &        &       &        &        &            &        &      &      &       &       &       & 2579 & 42.95 &      &       & 2813 & 22.17 \\
           &        &       &        &        &            &        &      &      &       &       &       & 2581 & 40.53 &      &       & 2819 & 22.54\\
           &        &       &        &        &            &        &      &      &       &       &       & 2584 & 35.70 &      &       & 2822 & 23.13 \\
           &        &       &        &        &            &        &      &      &       &       &       & 2586 & 32.54 &      &       & 2826 & 18.14 \\
           &        &       &        &        &            &        &      &      &       &       &       & 2588 & 30.31 &      &       & 2830 & 17.68 \\
           &        &       &        &        &            &        &      &      &       &       &       & 2591 & 26.58 &      &       & 2835 & 15.19 \\
           &        &       &        &        &            &        &      &      &       &       &       & 2593 & 21.01 &      &       & 2838 & 12.17 \\
           &        &       &        &        &            &        &      &      &       &       &       & 2595 & 18.03 &      &       & 2840 & 10.22 \\
           &        &       &        &        &            &        &      &      &       &       &       & 2598 & 17.11 &      &       & 2842 & 8.20 \\
           &        &       &        &        &            &        &      &      &       &       &       & 2600 & 13.76 &      &       & 2845 & 5.95 \\
           &        &       &        &        &            &        &      &      &       &       &       & 2602 &  9.11 &      &       & 2849 & 3.75 \\
           &        &       &        &        &            &        &      &      &       &       &       & 2605 &  7.81 &      &       & 2852 & 2.48 \\
           &        &       &        &        &            &        &      &      &       &       &       & 2607 &  6.51 &      &       & 2856 & 1.38 \\    
           &        &       &        &        &            &        &      &      &       &       &       & 2609 &  5.02 &      &       & 2867 & 0.62 \\   
           &        &       &        &        &            &        &      &      &       &       &       & 2616 &  3.16 &      &       & 2874 & 0.32 \\    
           &        &       &        &        &            &        &      &      &       &       &       & 2619 &  1.86 &      &       & 2882 & 0.18 \\     
           &        &       &        &        &            &        &      &      &       &       &       & 2633 &  0.50 &      &       & 2889 & 0.09 \\
           &        &       &        &        &            &        &      &      &       &       &       & 2656 &  0.33 &      &       & 2897 & 0.09  \\
           &        &       &        &        &            &        &      &      &       &       &       & 2685 &  0.15 &      &       &      &      \\
           &        &       &        &        &            &        &      &      &       &       &       & 2711 &  0.00 &      &       &      &      \\
\enddata
\end{deluxetable*}

\section{Results of Calibration}
\subsection{Photometric calibration}
All the calibration mentioned in Section 3 have been redone with more in-orbit 
observations. There are no surprises in the new results, but in many cases 
these show small but significant differences. 
\subsubsection{Zero point magnitudes}
%\begin{enumerate}
%\item 
%{\bf Zero point magnitudes:} 
Results of all the observations on HZ 4 have been 
used to recalculate zero point magnitudes for the filters. This magnitude is a 
measure of the sensitivity and its meaning is explained in the following 
sentences. Take a source which has a spectral shape identical to HZ 4 and which 
gives one count per second at the centre of the field, after applying 
all the corrections.
%correction for saturation (see Section 5.1.4). 
The average flux, within band of the filter, for this source would 
correspond to zero point magnitude at “$\lambda_{mean}$ (see 
Paper-1 for more details). 
For any filter, the observed count rates are first corrected to get equivalent 
"Normalised counts/sec" at the centre of the field by applying corrections for flat-field 
(including those reported below as “Remainders of flat field”), saturation (see section 5.1.4), and 
counts in the extended pedestal of PSF (see Table \ref{table-12}). The normalised counts/sec were 
used to calculate the zero point magnitude. To derive zero point (ZP) magnitude, 
an estimate is required for the average flux within band of the filter. In 
Paper-1, the average flux for HZ 4 was calculated within half 
power wavelengths of the filter, but here the full band of the filter (as shown 
in Table \ref{table-4}) is used for this calculation. The results are shown in Table \ref{table-3}. The 
errors shown correspond to 1$\sigma$ for the observed counts. Any errors related to 
correction for saturation and flat-field for any off-sets 
%from the centre are 
%expected to be no more than 1\%, except for N219M for which it could be 2\%.
from the centre are expected to be no more than 1\% for all the filters
except N219M for which these could be 2\%. 
%and are not included in the table.

%\item {\bf Effective areas:} 
\subsubsection{Effective areas}
The zero point magnitudes can be interpreted 
quantitatively only with reference to the spectral-shapes of the filters and the 
calibration source. Therefore, in order to fit any model spectral energy distribution to the observed 
c/s in the various filters, effective areas are required as a function of wavelength 
(see for example \citealt{2008MNRAS.383..627P}). These effective areas were 
calibrated on ground (see Paper-1) and are shown in Table \ref{table-4}.
%The in-orbit 
%calibrations of Hz4 provide a correction to these under the assumption that the 
%correction is independent of wavelength. 
The in-orbit calibration with HZ 4 were used to estimate corrections for the areas
under the assumption that relative change in transmission is independent of wavelength
for each filter.  These corrections are shown in Table \ref{table-5}. 
The corrected effective area is obtained by multiplying any entry in Table \ref{table-4} 
with the corresponding entry in Table \ref{table-5}. 
%The normalised counts/sec for HZ 4 shown in 
%Table \ref{table-3} and the effective areas provide all the observational inputs for 
%photometric calibration of UVIT. 

%\item {\bf Remainders of flat field:} 
\subsubsection{Remainders of flat field}
During the flat-field calibration on ground, 
only spatial variations in sensitivity of the detectors, at 1500 \AA ~for the FUV 
detector and at 2100 \AA ~for the NUV detector, were included and these were used 
for processing of the images. While the beam used for these calibration was 
expected to be uniform over small scales, e.g. over $\sim$10$^{\prime}$, it could 
have had variations over large scales. In addition, there could be spatial 
variations due to other optical elements, e.g. filters, and wavelength dependent 
variation in sensitivity of the detectors. In order to get a direct measure of 
the overall flat-field, exposures were taken for three fields in SMC. The first 
field was selected in a suitable part of SMC ($\alpha_{2000}$ = 01:09:46, 
$\delta_{2000}$ = $-$71:20:30). 
The second field was selected by applying a shift of $\sim$6$^{\prime}$ in one 
direction, and the third field was selected by applying the shift along an 
orthogonal direction.  These provided data for finding count rates, for a large 
number of sources, at three positions on the detector. The variations in these 
signals with position are called remainders 
of flat-field. Limited results on 
the remainders were reported in Paper-1.

%\end{enumerate}

\begin{deluxetable*}{llllllllll}
\tablecaption{The correction for the effective areas of the filters \label{table-5}.}
\tablewidth{0pt}
\tablehead{
\colhead{Filter}  & \multicolumn{9}{c}{Correction} 
}
%\decimalcolnumbers
\startdata
Filter     & F148W   & F154W & F169M & F172M & N242W & N219M & N245M & N263M & N279N  \\
Correction & 0.779   & 0.787 & 0.876 & 0.892 & 0.814 & 0.540 & 0.805 & 0.824 & 0.848 \\	 \hline
\enddata
\end{deluxetable*}

\begin{deluxetable*}{rrrrrrr}
\tablecaption{Parameters of the fit to the reminders for all the filters \label{table-6}. }
\tablewidth{0pt}
\tablehead{
\colhead{Parameter} &	\colhead{FUV-ALL}  &	\colhead{N242W}  & \colhead{N219M}  &	\colhead{N245M}	& \colhead{N263M} & 	\colhead{N279N} 
}
%\decimalcolnumbers
\startdata
a1  &	 3.15   $\times$ 10$^{-6}$  &	2.181$\times$10$^{-5}$    & -1.506 $\times$10$^{-5}$  &	 9.25 $\times$10$^{-6}$	 & 1.741$\times$10$^{-5}$  &  4.09$\times$10$^{-6}$   \\
a2  &	-2.879  $\times$ 10$^{-5}$  &  -1.55 $\times$10$^{-6}$	  &  1.85  $\times$10$^{-6}$  &	 1.14 $\times$10$^{-6}$	 &-5.46 $\times$10$^{-6}$  &  1.492$\times$10$^{-5}$  \\
a3  &	 3.00   $\times$ 10$^{-9}$  &	1.034$\times$10$^{-8}$	  &  9.541 $\times$10$^{-8}$  &	 1.379$\times$10$^{-8}$	 & 1.188$\times$10$^{-8}$  &  2.151$\times$10$^{-8}$  \\
a4  &	-2.51   $\times$ 10$^{-9}$  &	1.760$\times$10$^{-8}$	  &  6.761 $\times$10$^{-8}$  &	 1.188$\times$10$^{-8}$	 & 1.436$\times$10$^{-8}$  &  2.261$\times$10$^{-8}$  \\
a5  &	 3.30   $\times$ 10$^{-9}$  &	5.19 $\times$10$^{-9}$	  &  2.917 $\times$10$^{-8}$  &	 2.66 $\times$10$^{-9}$	 & 6.75 $\times$10$^{-9}$  &  1.517$\times$10$^{-8}$  \\
a6  &	-9.98   $\times$ 10$^{-12}$ &  -3.63 $\times$10$^{-12}$	  & -3.39  $\times$10$^{-12}$ &	 5.69 $\times$10$^{-13}$ &-4.46 $\times$10$^{-12}$ &  3.01$\times$10$^{-12}$  \\
a7  &	 1.232  $\times$ 10$^{-11}$ &	4.71 $\times$10$^{-12}$	  &  1.572 $\times$10$^{-11}$ &	 6.18 $\times$10$^{-12}$ & 1.103$\times$10$^{-11}$ &  1.159$\times$10$^{-11}$ \\
a8  &	 7.39   $\times$ 10$^{-12}$ &	3.86 $\times$10$^{-12}$	  &  2.186 $\times$10$^{-11}$ &	 3.45 $\times$10$^{-12}$ & 6.61 $\times$10$^{-12}$ &  8.33$\times$10$^{-12}$  \\
a9  &	-8.32   $\times$ 10$^{-12}$ &  -1.175$\times$10$^{-11}$	  &  1.750 $\times$10$^{-11}$ &	 1.95 $\times$10$^{-13}$ &-6.27 $\times$10$^{-12}$ & -1.96$\times$10$^{-12}$  \\
a10 &	 2.205  $\times$ 10$^{-5}$  &	9.905$\times$10$^{-5}$	  & -6.51  $\times$10$^{-6}$  &	 4.001$\times$10$^{-5}$	 & 2.899$\times$10$^{-5}$  &  3.885$\times$10$^{-5}$  \\
a11 &	-1.0635 $\times$ 10$^{-4}$  &  -2.54 $\times$10$^{-6}$	  &  1.835 $\times$10$^{-5}$  &	-5.29 $\times$10$^{-7}$	 &-2.468$\times$10$^{-5}$  &  1.664$\times$10$^{-5}$  \\
a12 &	-4.90   $\times$ 10$^{-6}$  &  -1.327$\times$10$^{-5}$	  &  6.826 $\times$10$^{-5}$  &	 2.87 $\times$10$^{-6}$	 & 4.98 $\times$10$^{-6}$  & -4.747$\times$10$^{-5}$  \\
a13 &	 4.03   $\times$ 10$^{-6}$  &	1.73 $\times$10$^{-6}$	  &  5.165 $\times$10$^{-5}$  &	 2.00 $\times$10$^{-6}$	 &-2.937$\times$10$^{-5}$  & -5.632$\times$10$^{-5}$  \\
a14 &	-6.772  $\times$ 10$^{-5}$  &	1.988$\times$10$^{-5}$	  &  3.2888$\times$10$^{-4}$  &	 3.837$\times$10$^{-5}$	 & 8.167$\times$10$^{-5}$ &  1.3243$\times$10$^{-4}$ \\
\enddata
\end{deluxetable*}

The flat-fields obtained on the ground are a good measure of small scale 
variations in sensitivity of the detectors. Therefore, the remainders are only 
expected over scales larger than $\sim$10 arcmin. The beam at the filters is 
$\sim$3.3 mm in diameter and any variations in the transmission on scales 
$<$ $\sim$1 mm are not of much consequence. In the ground calibration, all 
the filters except N219M gave $<$ 5\% peak-to-peak variations in transmission on scales 
$>$ 1 mm. Therefore we considered a third order polynomial in x \& y as a good 
choice to model the remainders, even if it misses some small scale variations 
for N219M.  In order to avoid errors related to drifting in and out of the 
sources near the edges, only data for sources falling within a radius of 1900 
sub-pixels were used, i.e. an extrapolation would be required to find remainders 
for radii $>$ 1900 sub-pixels. It was noted that a fit restricted to data for radii 
$<$ 1500 sub-pixels gave very small values for the remainders.  But a fit for all 
the radii gave much larger values for the remainders for radii $<$ $\sim$1500 
sub-pixels. This suggested that the relatively larger values at the outer parts of 
the field were biasing the fit for the central part. Therefore, the fit was made 
in two parts: (i) a third order polynomial for radii $<$ 1500 sub-pixels, and (ii) 
the change beyond radius of  1500 sub-pixels as linear  in radius with four 
parameters to generate the azimuthal dependence.  The function, $f$ a multiplicative factor, normalised to one 
at the centre, is written as

\begin{eqnarray}
f(x,y) = 1 + a_1x + a_2y + a_3x^2 +  ~~~~~~~~~~~~~~~~~~ \nonumber \\ 
a_4y^2 + a_5xy + a_6x^3 + a_7y^3 +  ~~~~~~~~~~ \nonumber \\
a_8yx^2 + a_9xy^2;  for R \le 1500 ~~~~~~~~~~  
\end{eqnarray}

\begin{eqnarray}
f(x,y) = 1 + a_1x + a_2y + ~~~~~~~~~~~~~~~~~~~~~~~~~~~~~~~~~~~~~~~~~ \nonumber \\
(1500/R)^2(a_3x^2 + a_4y^2 + a_5xy) + ~~~~~~~~~~~~~~~~~~~~\nonumber \\
(1500/R)^3(a_6x^3 + a_7y^3 + a_8yx^2 + a_9 xy^2) + ~~~~~~~~~~\nonumber \\
(R-1500) (a_{10}y/R + a_{11}x/R + a_{12}*2xy/R^2 + ~~~~~~~~~~\nonumber \\
a_{13} (x^2-y^2)/R^2 + a_{14}),  for R > 1500 ~~~~~~~~~~~~~~~~~~~~~~~~~                      
\end{eqnarray}

where x/y are the coordinates in sub-pixels as referred to the centre, and R is 
the radius.  As direct data on the sensitivity as a function of position were 
not available, an approximation was used. The observed count-rates for a source 
at different positions can be related to differentials in sensitivity as follows:   

\begin{equation}
(f(x1,y1) /f(x2,y2)) -1 = (s(x1,y1)/s(x2,y2)) - 1  %~~or                                                                 
\end{equation}
\begin{equation}
f(x1,y1)- f(x2,y2) = (s(x1,y1)/s(x2,y2) -1)*f(x2,y2)                                                    
\end{equation}
Here, s(x,y) is the sensitivity at x,y. As long as ``f'' does not differ much from 
unity , Equation 4 can be approximated as 

\begin{equation}
f(x1,y1) - f(x2,y2) \sim s(x1,y1)/s(x2,y2) -1                                                                         
\end{equation}

This procedure has three possible sources of error: (i) those due to finite 
statistics of the measured counts of individual objects and temporal variation 
of some sources (ii) those due to the approximation as per Equation 5, and (iii) due 
to any inadequacy in the choice of the function. Based on simulations, the errors 
due to the finite statistics are estimated to be $<$ 2\% within a radius of 
12$^{\prime}$ and $<$ 5\% for the full field, and errors due to the approximation 
are estimated to be $<$ 1\% for all the filters except N219M (the fitted 
values of ``$f$'' for all the filters except N219M range within 0.86 and 1.18). 
However, for N219M, the fitted values of ``$f$'' range within 1 and 1.5, and the errors 
would be $>>$ 1\%. Therefore, two iterations have been used for this filter 
where Equation 4 was used in the second iteration with the values of ``$f$'' obtained in 
the first iteration. The maximum values of ``$f$'' obtained in the first and second 
iterations are 1.41 and 1.49 respectively, and errors in the final values of ``$f$'' 
arising due to the approximation are estimated as $<$ 4\%. All the filters for 
FUV are crystaline and are not expected to deteriorate in orbit. The data for all 
the FUV filters were combined to get a common fit called FUV-ALL. The 
parameters obtained for all filters are given in Table \ref{table-6}. These fits were applied 
to the counts obtained for HZ 4, with F148W, N219M, and N279N, at multiple 
locations on the detectors. The results suggest that the remainders are corrected 
to better than 5\% for radii $<$ 12$^{\prime}$ ($\sim$1770 sub-pixels) on the detectors. This gives 
confidence in the process used. The actual fit was made for radii $<$ 1900 
sub-pixels, but the relations for radii $>$ 1500 sub-pixels can be used up to radii 
of 2000 sub-pixels. The number of source pairs used for the fits are: 
$\sim$8000 for FUV, $\sim$4000 for N242W, $\sim$1700 for N219M, $\sim$3100 for
N245M, and $\sim$1400 for N279N. Only those sources were included which had $>$ 
200 counts in at least one of the two fields. As maximum possible coverage was 
required over the area, all the sources were given equal weight irrespective of 
the counts. Images taken with filter N219M (NUV B15), which shows 
the largest variation in sensitivity across the field of view, were analysed to 
check the errors at radii between 1900 and  2000 sub-pixels.  A comparision of 
the corrected (as per the fit for radii $>$ 1500 sub-pixels) counts in images
of SMC1, SMC2, and SMC3 shows that the fractional rms errors are $\sim$0.06, a 
major part of which could be from Poisson statistics of the counts and errors of 
photometry near the edge.

\begin{figure*}
\vbox
    {
     \hbox{
     \hspace*{-0.5cm}\includegraphics[scale=0.40]{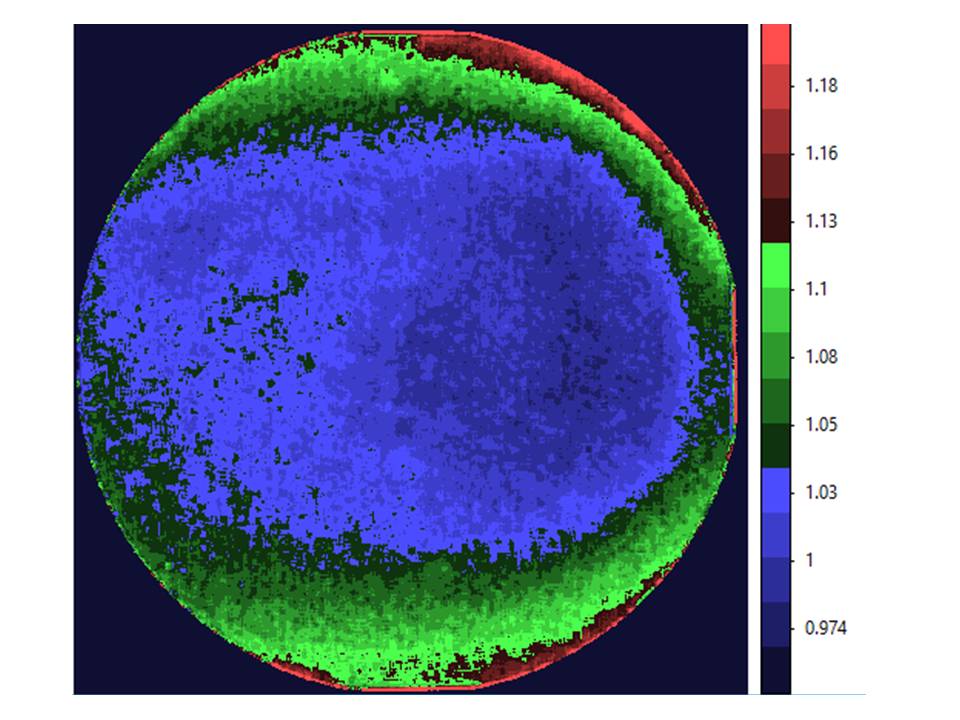}
     \hspace*{-0.5cm}\includegraphics[scale=0.40]{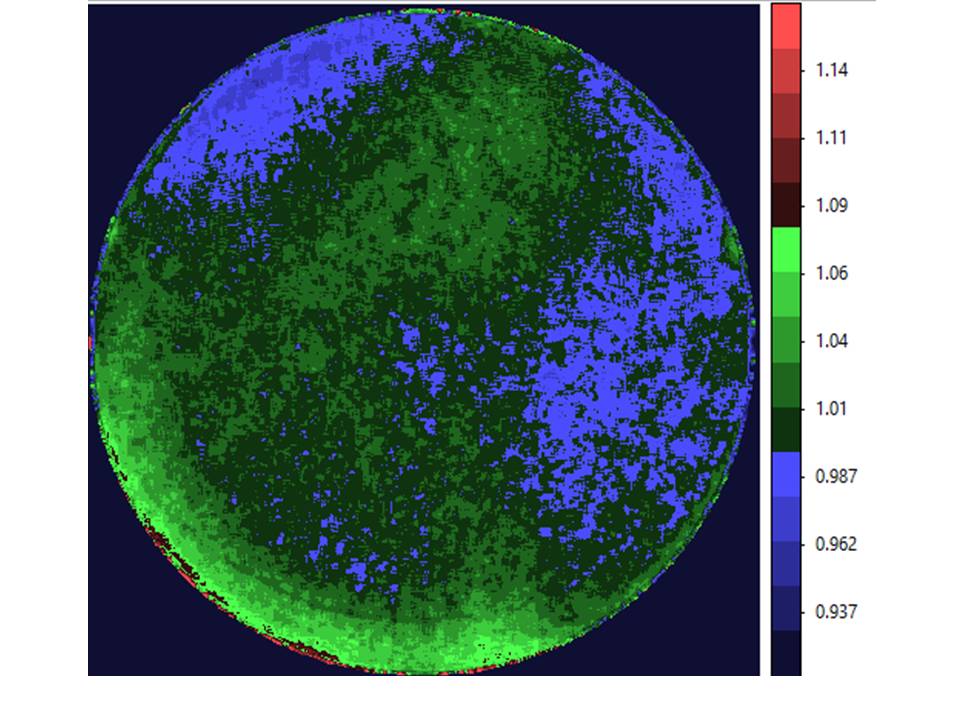}
          }
     \hbox{
     \hspace*{-0.5cm}\includegraphics[scale=0.40]{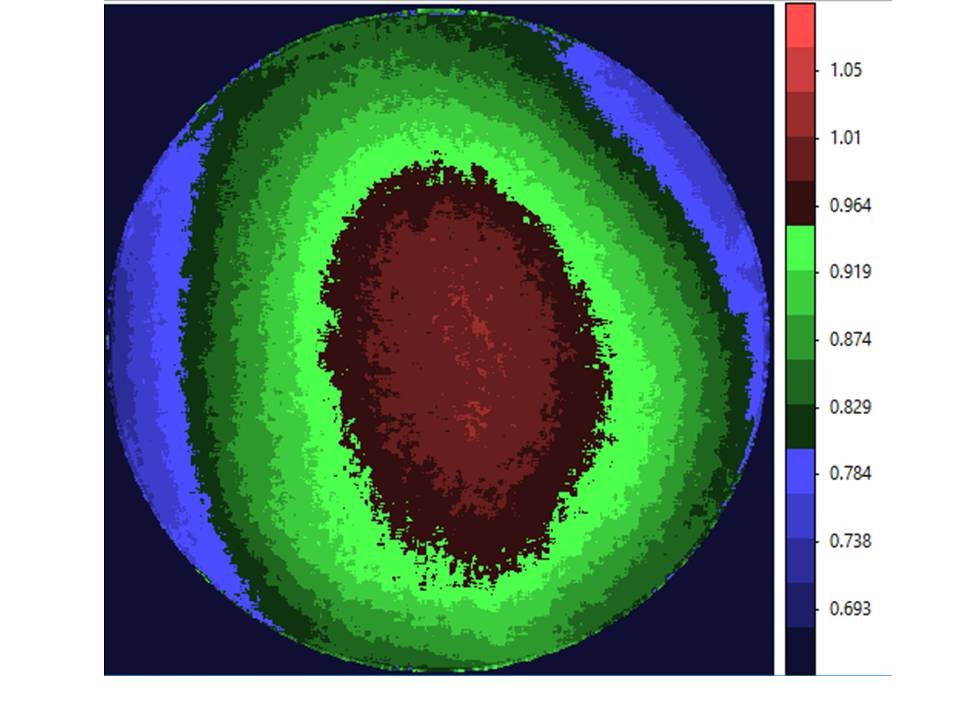}
     \hspace*{-0.5cm}\includegraphics[scale=0.40]{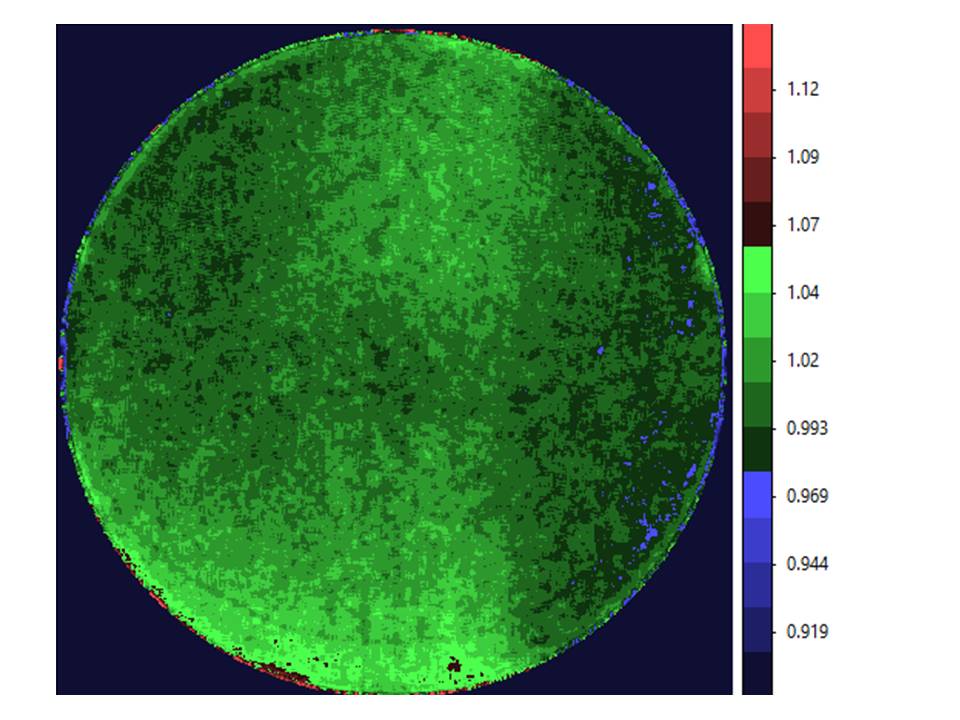}
          }
     \hbox{
     \hspace*{-0.5cm}\includegraphics[scale=0.40]{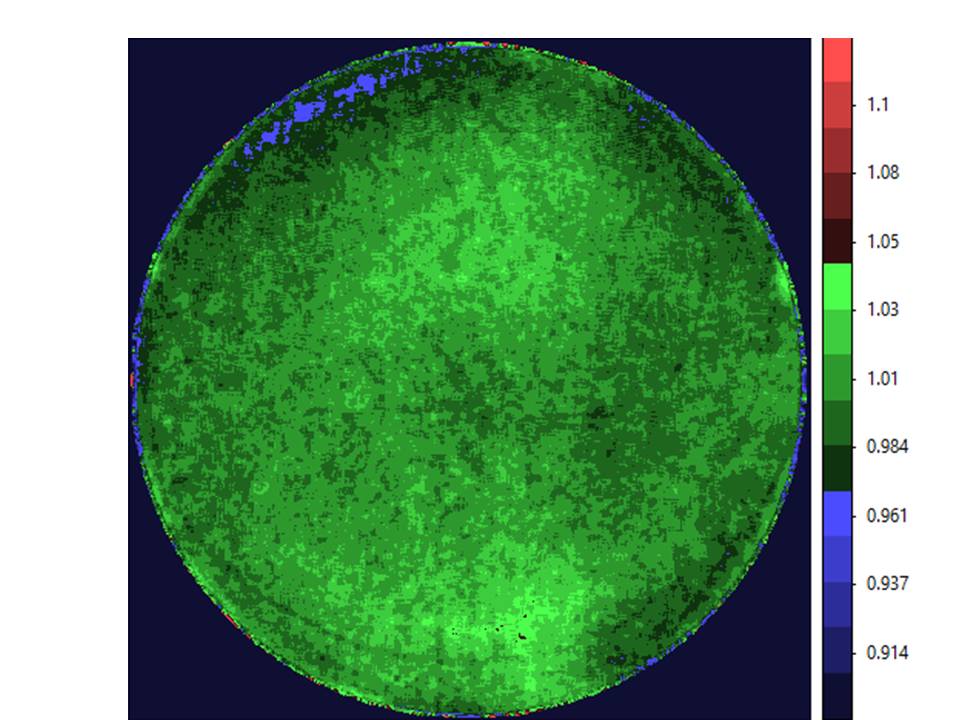}
     \hspace*{-0.5cm}\includegraphics[scale=0.40]{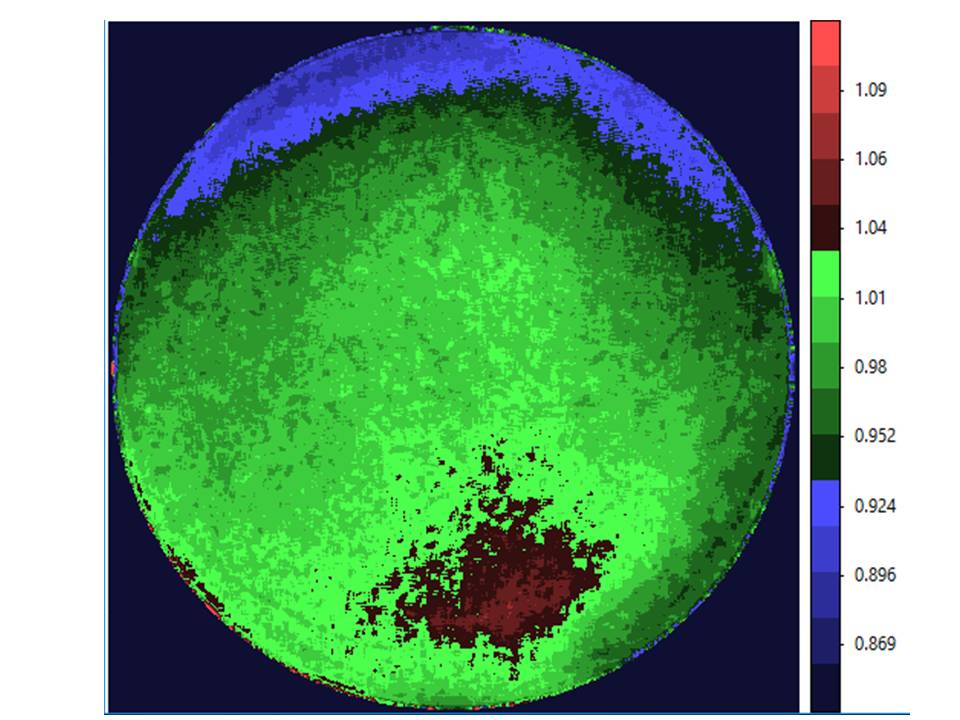}
          }
    }
\caption{Overall variations for inverse of sensitivity. 
The panels are: Top left: All FUV filters, Top right: N242W, 
Middle left: N219M, Middle right:  N245M, Bottom left: N263M , and Bottom right:  N279N \label{figure-1}}
\end{figure*}

Overall variations for inverse of the sensitivity are shown in Figure \ref{figure-1}. These 
variations were obtained by combining flat-fields of the detectors measured in 
the ground calibration with the results obtained here for remainders of 
flat-field, and were normalized to one at the centre.  We give in
Table \ref{table-7} two sets of  maximum and minimum of the inverse of
sensitivity over the field for different filters, one for radii $<$ 1900
sub-pixels and the other for radii $<$ 2000 sub-pixels. 

\begin{deluxetable}{lcccc}
\tablecaption{Range in the inverse of sensitivity for different filters 
\label{table-7}} 
\tablewidth{0pt}
\tablehead{
\colhead{Filter}       &\multicolumn{2}{c} {radii $<$ 1900 sub-pixels} &\multicolumn{2}{c} {radii $<$ 2000 sub-pixels}  \\
\colhead{}             & \colhead{Min.}  & \colhead{Max.}  & \colhead{Min.}  & \colhead{Max.}  
         }
%\decimalcolnumbers
\startdata
FUV(all filters) & 0.96       & 1.18       & 0.96       & 1.19 \\
N219M            & 0.72       & 1.03       & 0.68       & 1.03 \\
N242W            & 0.95       & 1.09       & 0.95       & 1.09 \\
N245M            & 0.96       & 1.08       & 0.96       & 1.08 \\
N263M            & 0.95       & 1.05       & 0.95       & 1.05 \\
N279N            & 0.88       & 1.07       & 0.86       & 1.08 \\
\enddata
\end{deluxetable}

\begin{deluxetable}{lrr}
\tablecaption{Log of observations of NGC 188 in the filters F148W and N279N\label{table-8}} 
\tablewidth{0pt}
\tablehead{
\colhead{}       &\multicolumn{2}{c} {Exposure Time (secs)} \\
\colhead{Date}   & \colhead{F148W}  & \colhead{N279N}    
         }
%\decimalcolnumbers
\startdata
31/12/2015  & 551.6  & 560.4  \\
13/07/2016  & ---    & 265.2  \\
30/01/2017  & 1194.7 & 1202.3 \\
16/04/2017  & ---    & 400.7  \\
21/12/2017  & 148.6  & 157.0  \\
22/02/2018  & 2893.5 & 1862.1 \\
04/04/2018  & 430.9  & ---    \\
20/07/2018  & 400.2  & ---    \\
26/08/2018  & 255.2  & ---    \\
13/09/2018  & 1163.3 & ---    \\
21/09/2018  & 1140.4 & ---    \\
23/10/2018  & 1140.3 & ---    \\
\enddata
\end{deluxetable}

\subsubsection{Correction for saturation}
In photon-counting mode, occurrence of multiple photon events in close 
proximity (within 3 $\times$ 3 pixels) in a frame is recorded as single photon. 
Therefore, there is some saturation unless the average photon rate per frame 
is $<<$ 1.  Process of correcting this remains the same as was reported 
in Paper-1, and it is described below.  

%{\bf We calculate the actual counts per frame (ICPF5), corresponding to
%97\% of the total counts per frame (CPF5), as per Poissonian relation
%between actual counts per frame and fraction of photonless frames.
%As the PSF is spread beyond 3 $\times$ 3 pixels, the actual 
%correction is less than this and is found as per the following
%equations.}

The actual counts per frame can be calculated 
by using relation of Poissonian statistics between 
the actual counts per frame and fraction of the frames
with no count. However, as the actual PSF is spread 
beyond 3 $\times$ 3 sub-pixels, we have to use an empirical procedure described below.
 Let “CPF5” be 97\% of the observed counts per frame and
let “ICPF5” be the corresponding actual counts per frame
as per Poissonian statistics. The equations 
used to get the correction for counts per frame are

\begin{equation}
CPF5 = (1 - exp(-ICPF5)) 
\end{equation}

\begin{equation}
ICORR = (ICPF5) - (CPF5) 
\end{equation}

\begin{equation}
RCORR = ICORR * (0.89 - 0.30 * (ICORR)^2) 
\end{equation}

where, ICORR is the ideal correction for saturation, RCORR is the real 
correction. It is found that 97\% of the total counts are contained within 
a window of radius $\sim$29 sub-pixels ($\sim$12$^{\prime\prime}$) which can be 
used to estimate “CPF5” in the above equations (see Section 5.4 on PSF). The 
parameters in the above equations are found empirically 
and it works well for point sources with observed (uncorrected) rates 
$<$$\sim$0.6 counts/frame. A similar correction for extended sources is more 
involved and is not discussed here.  As the observed frames for a point source
fall in two categories, i.e. either with one photon or
with no photon, the errors on actual counts are calculated as
per Binomial distribution.

\subsection{Reduction in Sensitivity of UVIT}
\begin{figure*}
\hspace*{-0.5cm}\includegraphics[scale=0.20]{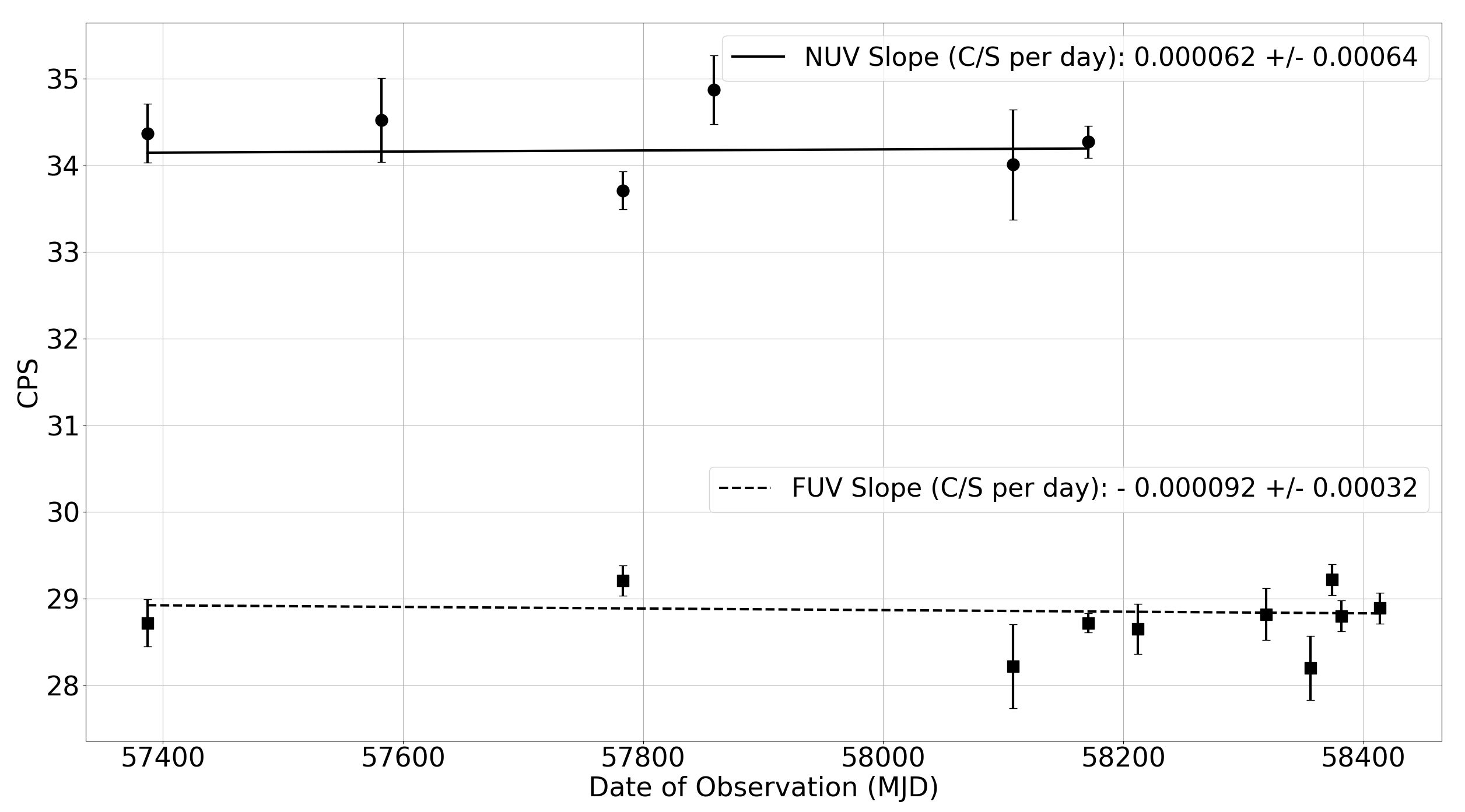}
\caption{Variations in the sensitivity of FUV and NUV channels of UVIT between December 2015 to October 2018 \label{figure-5}}
\end{figure*}

During all the manoeuvres, care was taken to keep bright earth 
and sun away from the field of view of UVIT, but scattered Ultra-Violet radiation from 
these sources could lead to slow deposition of contaminants \citep{1993SPIE.1971..276N} 
on the optics and hence could lead to a reduction in sensitivity of FUV and NUV channels.
In addition to this, factors such as the  ageing of the MCPs and filters, interaction of radiation with the
coatings on the optics could also lead to a reduction in the sensitivity of UVIT.
To estimate any degradation in the sensitivity of UVIT, 
signals for two stars in the field of NGC 188 were tracked between December 31, 2015 to 
October 23, 2018.  In FUV the observations were carried out in F148W filter,
while in NUV, the filter N279N was used. The log of observations of NGC 188 is given in Table \ref{table-8}.
In NUV, the two stars centered at ($\alpha_{2000}$ = 00:48:18.91, 
$\delta_{2000}$ = +85:13:26.04), 
($\alpha_{2000}$ = 00:42:43.68, $\delta_{2000}$ = +85:14:12.48) were used, 
while in FUV the two stars used 
for tracking the reduction in sensitivity are located at 
($\alpha_{2000}$ = 00:48:18.91, $\delta_{2000}$ = +85:13:26.04)  
and ($\alpha_{2000}$ = 00:47:52.15, $\delta_{2000}$ = +85:19:08.04) 
respectively.
The average saturation corrected counts/sec (obtained over a circular 
aperture of 50 sub-pixels radius, that encompasses more than 98\% of the energy 
in both FUV and NUV) of the two stars in FUV and NUV are given in 
Fig. \ref{figure-5}. Our results are
consistent with no 
reduction in the 
sensitivity of FUV and NUV channels of UVIT. 

\subsection{Spectroscopic Calibration}
Spectroscopic calibration has been done with observations of NGC40 (for 
dispersion) and HZ 4 (for effective areas). The data have been reanalysed by 
\cite{gulab}. The results of this analysis are presented here. 
For typical images with the gratings and other details please refer to 
Paper-1. We note that in Paper-1: 
(i) the "pixels" are twice in linear size as compared to the sub-pixels used here, 
(ii) the nomenclature for the gratings FUV1 and FUV2 is inverted relative to the nomenclature used 
here.

\subsubsection{Dispersion and Spectral Resolution}
\begin{figure}
\hspace*{-0.5cm}\includegraphics[scale=0.36]{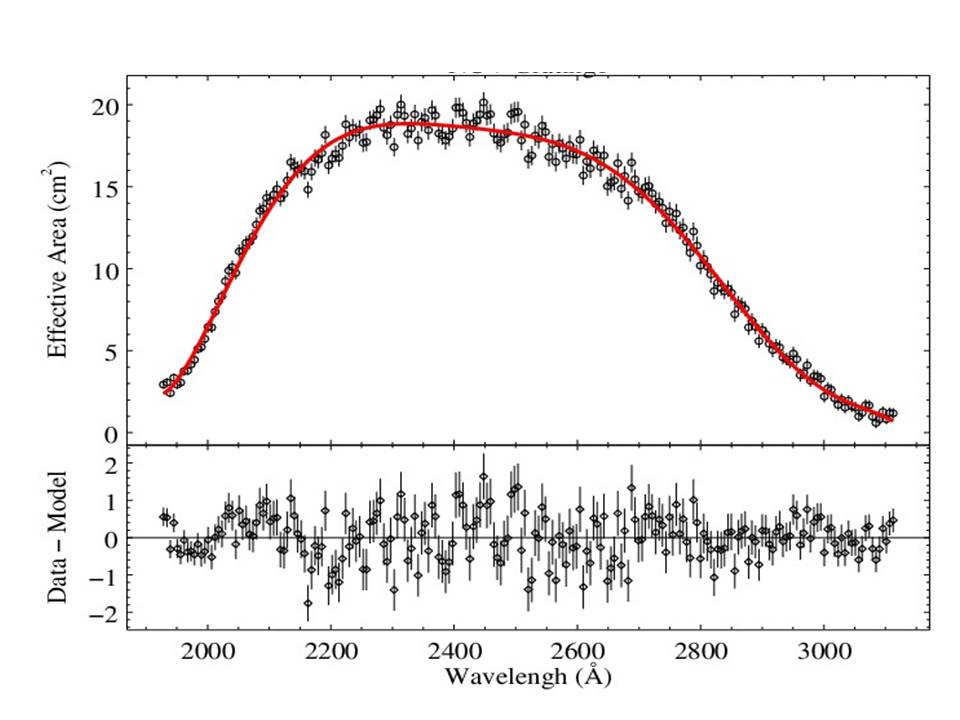}
\caption{Effective area of the NUV-grating as a function of wavelength \label{figure-2}}
\end{figure}

\begin{figure}
\hspace*{-0.5cm}\includegraphics[scale=0.36]{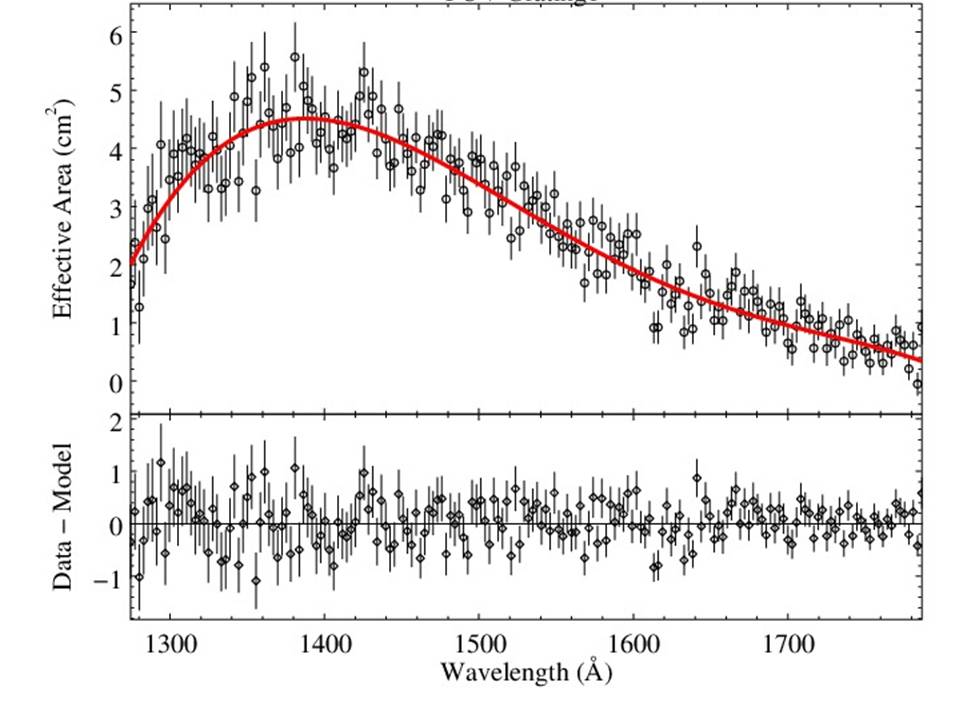}
\caption{Effective area of the FUV1-grating as a function of wavelength \label{figure-3}}
\end{figure}

\begin{figure}
\hspace*{-0.5cm}\includegraphics[scale=0.36]{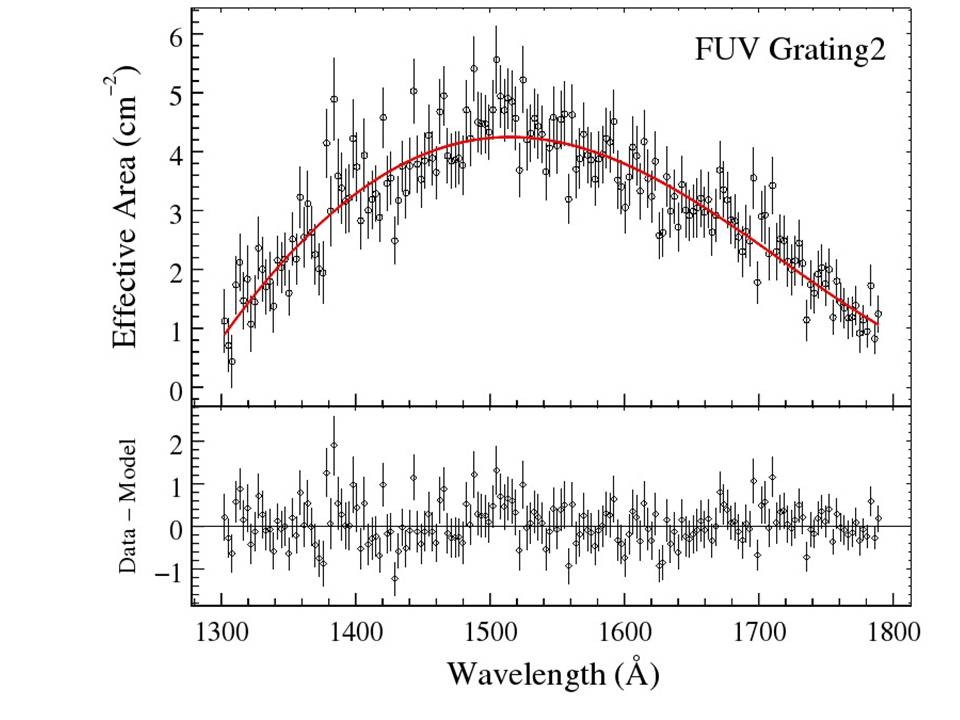}
\caption{Effective area of the FUV2-grating as a function of wavelength \label{figure-4}}
\end{figure}

The relations between wavelength (in Angstroms) and shift (in sub-pixels) 
from zero order, and spectral resolutions, are shown in Table \ref{table-9}.
Here shifts (in sub-pixels) are written as x or y to indicate whether the 
dispersion is along rows/columns.  
The effective areas plots are shown in Figure \ref{figure-2} to Figure \ref{figure-4}, 
and the polynomial fits to the effective areas are given in Table \ref{table-10}.

%\subsubsection{Effective Areas}

\begin{deluxetable}{llc}
\tablecaption{Dispersion solution for the different gratings of UVIT \label{table-9}. }
\tablewidth{0pt}
\tablehead{
%\colhead{NUV first order} &	\colhead{FUV1 Second order}  &	\colhead{FUV2 second order}  
\colhead{Order} &	\colhead{Equation}  &	\colhead{Spectral resolution}  
}
%\decimalcolnumbers
\startdata
NUV first order & $\lambda (\AA) = -5.5868 x + 18.1$ & 38  \AA \\
FUV1 second order & $\lambda (\AA) = -2.799 x + 40.2$  & 16 \AA \\
FUV2 second order & $\lambda(\AA) = -2.813 y + 33.9$   & 14 \AA \\
%$\lambda (\AA) = -5.5868 x + 18.1$ & $\lambda (\AA) = -2.799 x + 40.2$  &  $\lambda(\AA) = -2.813 y + 33.9$ \\
%Spectral resolution = 38 \AA       & Spectral resolution = 16 \AA       & Spectral resolution = 14 \AA  \\
\enddata
\end{deluxetable}

\begin{deluxetable*}{ll}
\tablecaption{Polynomial fits to the effective areas (EA) for
the gratings \label{table-10}. }
\tablewidth{0pt}
\tablehead{
%\colhead{NUV first order} &	\colhead{FUV1 Second order}  &	\colhead{FUV2 second order}  
\colhead{Parameter} &	\colhead{Value}    
}
%\decimalcolnumbers
\startdata
NUV first order  & EA(cm$^2$) = 900363.87 -  2548.8671167 $\lambda$ + 3.07510804796 $\lambda^2$  -  \\
                 & ~~~~~~~~~~~~~~~~~~0.0020498923174 $\lambda^3$ + 8.1553032340 * 10$^{-7}$ $\lambda^4$ - 1.93655027784*10$^{-10}$ $\lambda^5$ + \\
                 & ~~~~~~~~~~~~~~~~~~2.5415821036*10$^{-14}$ $\lambda^6$ -  1.42229434666*10$^{-18}$ $\lambda^7$ \\
FUV1 second order & EA (cm$^2$)= -3394.60 +  8.504523 $\lambda$ -  0.0079305062 $\lambda^2$ + 3.2687397*10$^{-6}$ $\lambda^3$ - 5.031413*10$^{-10}$ $\lambda^4$  \\
FUV2 second order & EA (cm$^2$)= 268.14 - 1.033632 $\lambda$ +  0.0012895741 $\lambda^2$ - 6.5494929*10$^{-7}$ $\lambda^3$  +1.1761863*10$^{-10}$ $\lambda^4$ \\
\enddata
\end{deluxetable*}

%EA(cm$^2$) = 900363.87 -                  & EA (cm$^2$)= -3394.60 +              & EA (cm2)= 268.14 -                  \\
%2548.8671167 $\lambda$ +                  & 8.504523 $\lambda$ -                 & 1.033632$\lambda$ +                 \\
%3.07510804796 $\lambda^2$  -              & 0.0079305062 $\lambda^2$ +           & 0.0012895741 $\lambda^2$ -          \\
%0.0020498923174 $\lambda^3$ +             & 3.2687397*10$^{-6}$ $\lambda^3$ -    & 6.5494929*10$^{-7}$ $\lambda^3$     \\
%8.1553032340 * 10$^{-7}$ $\lambda^4$ -    & 5.031413*10$^{-10}$ $\lambda^4$      & +1.1761863*10$^{-10}$ $\lambda^4$   \\
%1.93655027784*10$^{-10}$ $\lambda^5$ +   &                                      &                                     \\
%2.5415821036*10$^{-14}$ $\lambda^6$ -    &                                      &                                     \\
%1.42229434666 *10$^{-18}$                 &                                      &                                      \\
%\enddata
%\end{deluxetable}

\subsection{Point Spread Function (PSF) Calibration}
\begin{figure}
\hspace*{-0.5cm}\includegraphics[scale=0.36]{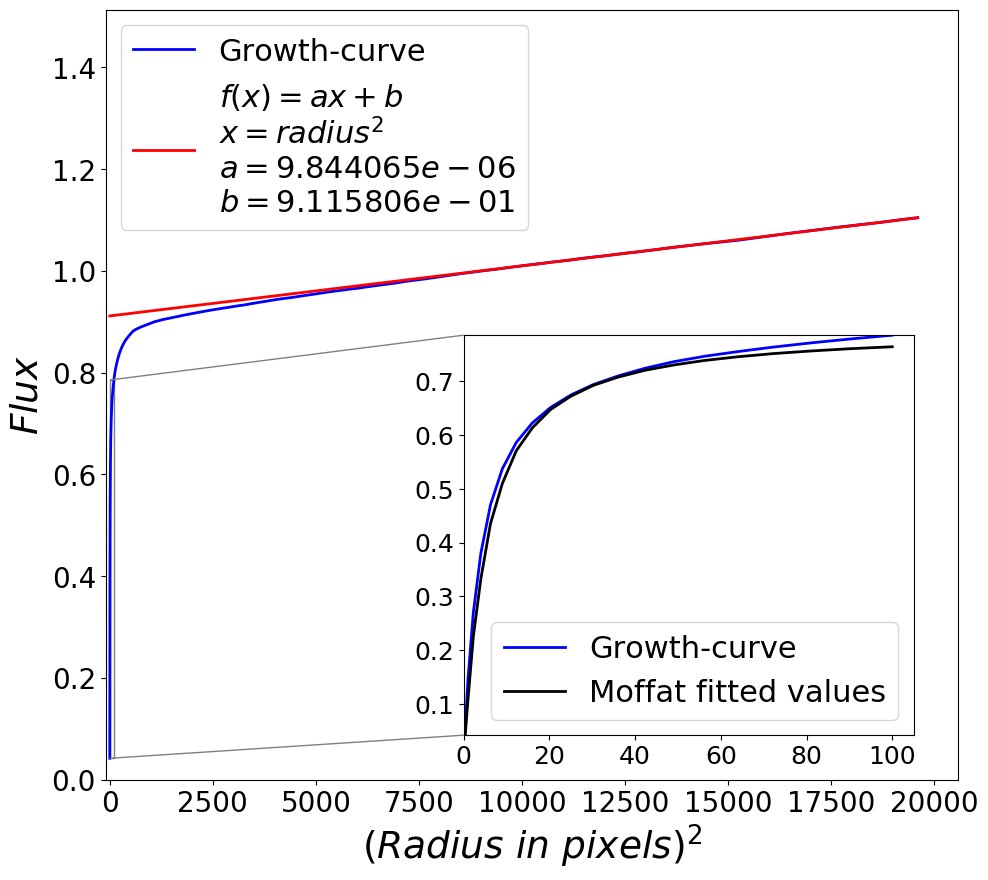}
\caption{Growth curve (normalised by the flux obtained from 95 sub-pixels radius) 
for the PSF for NUV with N242W filter. The 
Y-axis shows total counts seen in a circle for which square of the radius 
(in sub-pixels) is shown on X-axis. (Small inset on top-left shows equation 
for the fit, and large inset at bottom-right shows magnified view of the curve.) \label{figure-6}  }
\end{figure}

\begin{figure}
\hspace*{-0.5cm}\includegraphics[scale=0.36]{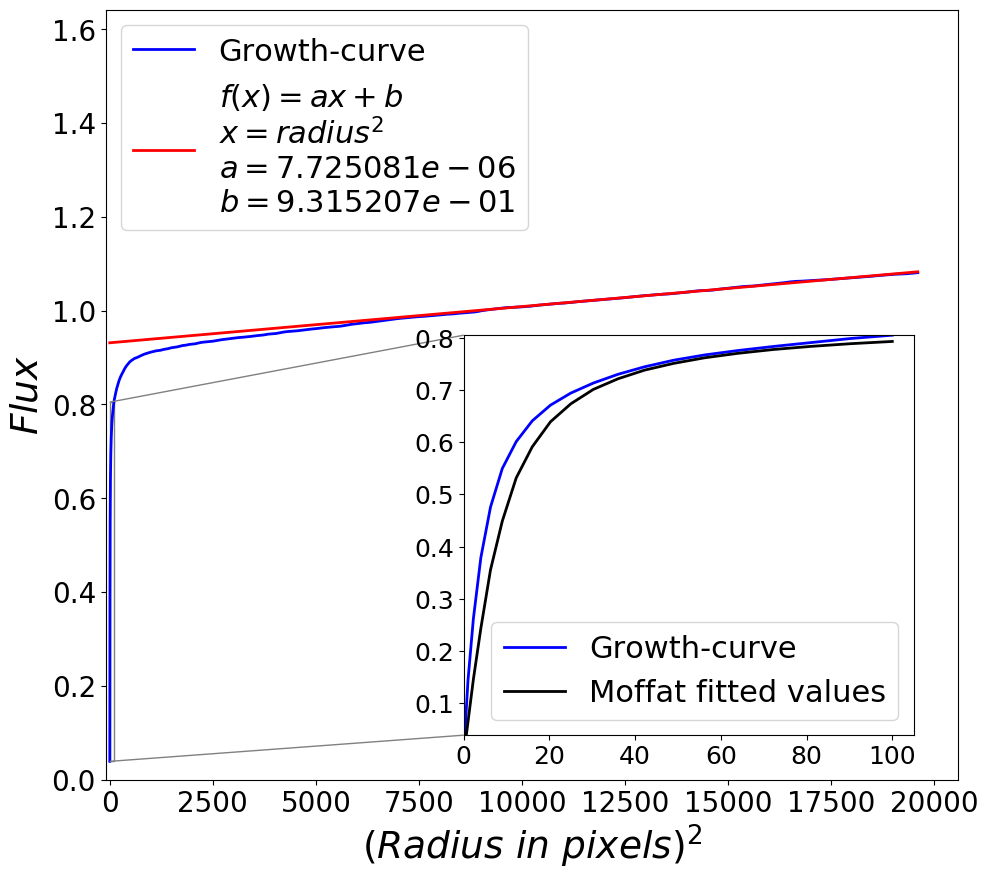}
\caption{Growth curve (normalised by the flux obtained from 95 sub-pixels radius) 
for the PSF  for FUV with F148W filter. The Y-axis 
shows total counts seen in a circle for which square of the radius 
(in sub-pixels) is shown on X-axis. (Small inset on top-left shows equation for 
the fit, and large inset at bottom-right shows magnified view of the curve.) \label{figure-7}}

\end{figure}

The PSF consists of a narrow core and an extended pedestal. New results on PSF
 were obtained with observations of HZ 4 with short exposure frames to minimise 
the effect of saturation. In the photon counting mode, frames for individual 
exposures are analysed by the on-board hardware to communicate position of the 
detected photons. For shorter exposures (larger read rates) the probability of two photons occurring 
in one frame is less. Correction for saturation was applied as per Section 5.1.4, and 
it was assumed that the correction is uniform and is limited to a circle of 
radius 22.5 sub-pixels of the PSF. The results show that the pedestal is much 
more extended as compared to what was reported in Paper-1 and contains $\sim$3\% 
more energy. The growth curves for NUV and FUV are shown in Figure \ref{figure-6}  
and Figure \ref{figure-7} and Table \ref{table-11}.

The pedestal is caused by scattering due to mechanical blocks and 
roughness/aberrations of the mirrors and filters and its strength is not 
expected to vary significantly with position on the detector.   
The observed variations in fractional energy of the pedestal for different 
filters are shown in Table \ref{table-12}. 
%From these variations it seems that the filters 
%contribute significantly to the scattering. Further, silica filters (F172M 
%and N242W) seem to scatter the least while the multilayer-interference filters 
%(N219M, N245M, N263M, and N279N) seem to scatter more. 
These data indicate that the multilayer-interference filters scatter more than the
crystalline filters.

Table \ref{table-13} shows that FWHM varies from $\sim$2.1 sub-pixels 
($\sim$0.9$^{\prime\prime}$) to $\sim$3.4 sub-pixels ($\sim$1.4$^{\prime\prime}$) 
for radii $\le$ 12$^{\prime}$ (fits to 2-D symmetric Gaussian give ~ 15\% larger 
values for FWHM). As reported in Paper-1, for larger radii ($>$12$^{\prime}$) 
the 
FWHM could increase up to $\sim$2$^{\prime\prime}$ due to the distortions. We 
note that FWHM can be adversely affected for individual exposures due to larger 
perturbations in the pointing or defocus due to any large variation in the 
temperature. Ground calibration data acquired by changing the temperature
of $\pm$5 deg around the operating temperature of 20 deg found neglible changes in
the telescope focus. The temperature stability achieved in space is much lesser
than $\pm$ 3 deg \citep{sriram}. From repeated observations of NGC 188, we have not
found any change in the PSF with time. An analysis of the PSFs for non-saturated images shows that
the sharpness of the NUV-PSF is underestimated for radii $<$7 sub-pixels; this
seems to be due to a larger saturation for NUV Silica (N242W) image of HZ4
($>$ 0.4 c/f) and the assumption that saturation is uniform for
radii $<$ 22.5 sub-pixels.

Possible causes for variation in FWHM of the core are (a) 
defocus, due to curvature of the focal plane, tilt of the detector-plane, 
dispersion etc., (b) variations in errors in tracking of the drift, and (c) 
variations in intrinsic spatial resolution of the detector with wavelength. 
Images of a part of SMC were used to find FWHM, by fitting a symmetric Moffat 
profile, for a large number of stars within a radius of 12$^{\prime}$’.  The results are 
shown in Table \ref{table-13}. Some systematic trends can be inferred from these data: (a) 
except for the filter F148W, focus for all the other FUV-filters is not optimal 
and the effects of curved focal plane are evident, (b) for the NUV filters the 
focus is optimal and the effects of curved focal plane are not significant, (c) 
for the FUV-filters, dispersion seems to broaden the FWHM, (d) for the NUV-filters 
the FWHM are better as compared to those for FUV-filters $-$ this could result 
from a combination of less dispersion in NUV and  a better spatial resolution 
of the NUV-detector, and (e) filters for longer wavelengths show better FWHM $-$ 
this is most likely due to better spatial resolution of the detector at 
wavelengths closer to the cut-off wavelength resulting from lower lateral 
momentum/movement of the photo-electrons between the photo-cathode and MCP 
(see  \citealt{2006SPIE.6294E..0WB} for more details on this effect).  

\begin{deluxetable}{lll}
\tablecaption{Encircled energy as a function of radius in sub-pixels. This
is based on analysis of the data obtained using Silica filter in NUV and 
CaF2 filter in FUV \label{table-11}.}
\tablewidth{0pt}
\tablehead{
\colhead{Radius}  &   \colhead{\% Energy (NUV)}  & \colhead{\% Energy (FUV)} 
}
%\decimalcolnumbers
\startdata
1.5     &  29.9      & 28.1 \\
2.0     & 42.0       & 40.7 \\
2.5     & 52.0       & 51.1 \\
3.0     & 59.3       & 59.1 \\
4.0     & 68.8       & 68.9 \\
5.0     & 74.5       & 74.6 \\
7.0     & 81.3       & 81.4 \\
9.0     & 85.1       & 85.0 \\
12.0    & 89.3       & 88.6 \\
15.0    & 92.1       & 91.3 \\
20.0    & 95.2       & 94.5 \\
30.0    & 97.6       & 96.9 \\
40.0    & 98.4       & 97.7 \\
50.0    & 98.8       & 98.3 \\
70.0    & 99.4       & 99.1 \\
80.0    & 99.6       & 99.5 \\
95.0    & 100.0      & 100.0 \\
\enddata
\end{deluxetable}

\begin{deluxetable*}{llllllllll}
\tablecaption{Observed variations of energy (percentage of the total) in the pedestal. 
Energy in the pedestal is defined as energy between radii of 7 
sub-pixels and 100 sub-pixels \label{table-12}.}
\tablewidth{0pt}
\tablehead{
\colhead{Filter}   & \colhead{F148W} & \colhead{F154W} & \colhead{F169M} & \colhead{F172M} & \colhead{N242W} &\colhead{N219M} &\colhead{N245M} 
&\colhead{N263M} & \colhead{N279N}
}
%\decimalcolnumbers
\startdata
\% of energy & 18.6 $\pm$ 0.3    & 18.2 $\pm$ 0.3  & 18.2 $\pm$ 0.3 & 17.2 $\pm$ 0.3   & 18.7 $\pm$ 0.3  & 20.9 $\pm$ 
0.6   & 19.6 $\pm$ 0.3 & 19.6 $\pm$ 0.3   & 21.8 $\pm$  0.6 \\ 
\enddata
\end{deluxetable*}

%\begin{figure}
%\hspace*{-0.5cm}\includegraphics[scale=0.40]{astromeric_diffFUV_NUV_new.jpg}
%\caption{:  Relative differences in positions (in sub-pixels) of the stars in 
%the UVIT images taken with FUV-F154W and NUV-N263M are shown as vectors, where 
%tail of the vector corresponds to position in the NUV field. For radii between 
%1500 and 1800 sub-pixel (10$^{\prime}$ to 12$^{\prime}$), the rms difference 
%is $<$ 0.9 sub-pixels ($\le$ 0.4$^{\prime\prime}$) and it increases to 
%1.7 sub-pixels (0.7$^{\prime\prime}$) for radii $>$ 1900 
%sub-pixel ($\sim$ 13$^{\prime}$)\label{figure-8}.}
%\end{figure}

\begin{figure}
%\hspace*{-0.5cm}\includegraphics[scale=0.40]{Astrometric_difference_FUV_NUV_old.jpg}
\hspace*{-0.5cm}\includegraphics[scale=0.40]{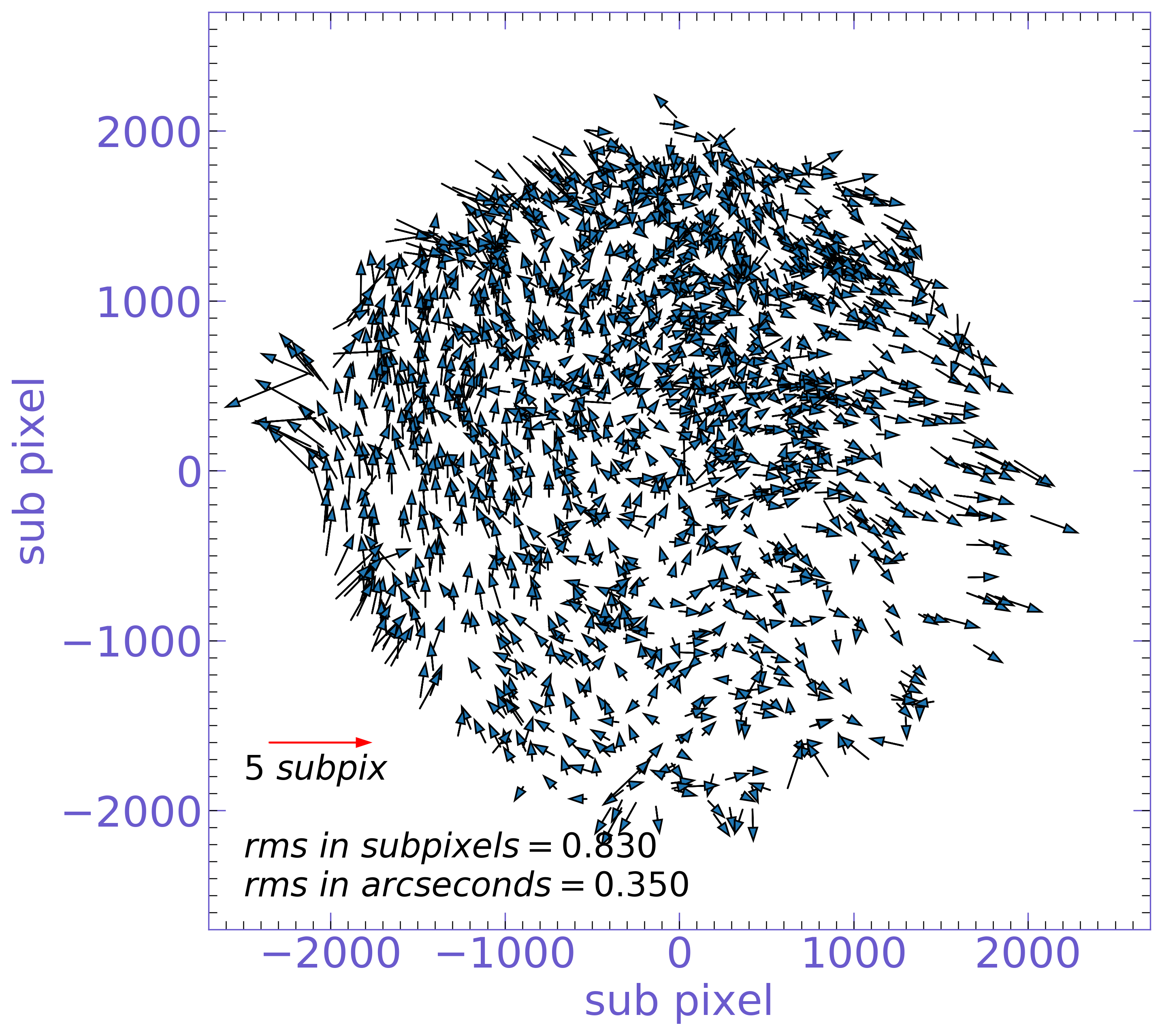}
\caption{Relative differences in positions of the stars in the UVIT images of SMC taken 
with FUV-F154W and NUV-N263M are shown as vectors, where tail of the vector 
corresponds to position in the NUV field. Positions and errors are shown in 
sub-pixels. For radii $<$ 1900 sub-pixels, any source giving an error $>$ 2 sub-pixels
is either a close double or is at a radius $>$ 1900 sub-pixels in the FUV detector \label{figure-8}}
\end{figure}

\subsection{Astrometric Calibration}
Positions measured by the detectors show deviations from linearity, i.e. the 
detectors show distortion. The distortions were calibrated on the ground and 
a part of these data were used to correct the measured positions for the 
results reported in Paper-1. Now, all the data on calibration have been  used to correct the 
measured positions. The resulting corrections differ significantly for NUV 
near edges in the first quadrant. To get an estimate of the distortion in
the final images, positions of stars in the NUV-images of SMC  with filter N263M were 
compared with the positions in the FUV image  with filter F154W, after due 
corrections for the relative plate scale and shift \& rotation between the two images. 
%The differences are shown in Figure \ref{figure-8} and for comparision the 
%results reported earlier are shown in Figure \ref{figure-9}. 
%A comparison of the two figures 
%shows that for $x$ values  $>$ 1800 sub-pixel in the first quadrant, the new results are significantly 
%better. 
The differences are shown in Figure \ref{figure-8}. It is seen that the new results
are significantly better than those reported in Paper-1 for x $>$ 1800 sub-pixels
in the first quadrant.
Averaged over the full field, rms of the deviations is $<$ 0.4$^{\prime\prime}$. 
while it is 0.3$^{\prime\prime}$ within a diameter of 24$^{\prime}$.
As the optics and the detectors for FUV and NUV are independent, these  
differences can be taken as upper bound for leftover distortion in the individual images. 

\begin{deluxetable*}{llllllllll}
\tablecaption{Observed variations (in sub-pixels)  of the FWHM with positions across the detector. For FUV filters, 
the first number is for stars at radii $<$  7.5 arcmin, and the second number is for 
stars at radii $>$ 7.5 armimn and $<$ 12 arcmin, while for NUV filters the number is for all the stars within a radius of 12 arcmin and
no significant variation in FWHM with radius was seen \label{table-13}.}
\tablehead{
\colhead{Filter}   & \colhead{F148W} & \colhead{F154W} & \colhead{F169M} & \colhead{F172M} & \colhead{N242W} &\colhead{N219M} &\colhead{N245M} 
&\colhead{N263M} & \colhead{N279N}
}
%\decimalcolnumbers
\startdata
FWHM      &   3.06,3.11  & 3.37,3.11 & 3.10,2.66 & 3.15,2.78 & 2.63     & 2.75   & 2.59  & 2.10  & 2.19  \\ 
\enddata
\end{deluxetable*}

\section{Conclusion}
Based on additional data, results on the calibration for UVIT have been revisited. Results on all calibration
except flat field show only small differences as compared to the results 
reported in Paper-1. Details on the new calibration that are needed for the
users of UVIT data along with the updated CALDB are also available at https://uvit.iiap.res.in/. We summarize here the findings 
of the calibration 

\begin{enumerate}
\item Photometric calibration: (a)as compared to estimates from the ground 
calibration, reduction in sensitivity ( or average effective area) for all the 
filters except N219M lie in range of 11\% to 22\%, while the filter N219M shows a 
reduction of $\sim$46\%. These differences are most likely related to inaccuracies
in the ground calibration which were only meant to assure that the sensitivities
in different bands were not less than 50\% of the designed value. (b) The peak to peak variation in sensitivity with position within a diameter of 
24$^{\prime}$ is $<$ 20\% for all the filters except N219N for which it is 
$\sim$35\% and (c) new zero point magnitudes are estimated with reference to the white dwarf HZ 4.
\item Repeated observations of  2 stars in NGC 188 over 3 years in the orbit show 
no reduction in the sensitivity of the FUV and NUV channels. 
\item Spectroscopic calibration give a peak effective area of 
$\sim$18.5 sq cm and a resolution of 38 \AA ~for the NUV grating, and a peak 
effective area of  $\sim$4.5 sq cm and a resolution of $\sim$15 \AA ~for the FUV gratings,
\item The PSF shows a FWHM  1.4$^{\prime\prime}$ or better within a 
diameter of 24$^{\prime}$ for FUV as well as NUV.
\item Astrometric calibration in orbit shows that the uncorrected distortion  
is $<$ 0.3$^{\prime\prime}$  rms within a diameter of  24$^{\prime}$.
\end{enumerate}

\acknowledgements

The UVIT project is a result of collaboration between IIA, Bengaluru, IUCAA, 
Pune, TIFR, Mumbai, several centers of ISRO, and CSA. Several groups from ISAC 
(ISRO), Bengaluru, and IISU(ISRO), Trivandrum have contributed to the design, 
fabrication,and testing of the payload. The Mission Group (ISAC)and ISTRSAC 
(ISAC) of ISRO have provided support in making the observations, and reception 
and initial processing of the data. We gratefully thank all the members of 
various teams for providing support to the project from the early stages of 
design to launch and observations in the orbit. Help by Shree Akash Vani in 
comparing the observations on HZ 4 and NGC 188 with the new flat-field corrections 
is gratefully acknowledged. 

\bibliography{ref.bib}{}
\bibliographystyle{aasjournal}

\end{document}